# Context-Augmented Code Generation Using Programming Knowledge Graphs

Iman Saberi [1]   Fatemeh Fard [1]

## Abstract

Large Language Models (LLMs) and Code-LLMs (CLLMs) excel at code generation but struggle with complex problems. Retrieval-Augmented Generation (RAG) mitigates this by integrating external knowledge, yet retrieval models often miss relevant context, and generation models hallucinate with irrelevant data. We propose a framework leveraging a Programming Knowledge Graph (PKG) for semantic representation and fine-grained retrieval of code and text. Our approach enhances retrieval precision through tree pruning and reduces hallucinations via a re-ranking mechanism that integrates non-RAG solutions. Structuring external data into finer-grained nodes improves retrieval granularity. Evaluations on HumanEval and MBPP show up to 20% pass@1 accuracy gains and a 34% improvement over baselines on MBPP. Key contributions include PKG-based retrieval, tree pruning, and robust re-ranking.

## 1. Introduction

Large Language Models (LLMs) have significantly improved the performance of tasks related to code, such as code generation (Huang et al., 2023; Roziere et al., 2023a; Li et al., 2023; Wang et al., 2023). As code-related models continue to emerge rapidly (Chen et al., 2021; Li et al., 2023; 2022; Roziere et al., 2023a; Zhu et al., 2024), most of these models rely on a natural language-to-code (NL-to-Code) paradigm, which often lacks the ability to leverage existing contextual information (Wang et al., 2024). Generating a solution from scratch, without access to supplementary context, poses significant challenges (Wang et al., 2024), even for humans (Zhong et al., 2024).

[1]Department of Computer Science, University of British Columbia Okanagan, Kelowna, Canada. Correspondence to: Iman Saberi <iman.saberi@ubc.ca>, Fatemeh Fard <fatemeh.fard@ubc.ca>.

Retrieval-Augmented Generation (RAG) enables retrieving and integrating relevant context from external knowledge sources during the inference time (Guu et al., 2020; Lewis et al., 2020), minimizing the necessity of embedding all knowledge within the model's parameters (Asai et al., 2024). RAG-based approaches can enhance accuracy across different scenarios (Izacard et al., 2022), without the need for further training of the model (Mallen et al., 2022; Ram et al., 2023). RAG-methods for code generation were previously proposed for retrieving information from library documentation (Zhou et al., 2022a) and file repositories (Zhang et al., 2023). Wang et al. (2024) explored the impact of different retrieved chunk sizes or including the entire data cells during the retrieval for code generation; showing that both factors have a negative effect on the performance of code generation tasks by introducing irrelevant data. They identified two main challenges in retrieval for code generation. First, accurately identifying and retrieving helpful documents, and second, the limited context capacity of models that can lead to hallucinations when given irrelevant data. Our work aims to alleviate these challenges through two main contributions.

To retrieve accurate data, we propose **Programming Knowledge Graph (PKG)** to represent source code or code-related corpus. Each node in PKG represents a code block extracted from function's Context-Flow Graph (CFG) or a code concept extracted from a Directed Acyclic Graph (DAG) corresponding to a textual chunk. PKG supports enabling an effective semantic search to retrieve the best-matching node given a query. We then apply **tree pruning** to remove irrelevant branches, ensuring that only the most useful information is passed to the generative model.

To address the second challenge, we propose a **re-ranker** model that combines outputs from multiple methods (e.g., RAG and non-RAG approaches) and re-ranks the generated solutions. As shown in Figure 1, different approaches excel at solving distinct types of problems, demonstrating the need for a re-ranker. When the initial retrieved content introduces hallucinations into the output, the re-ranker can prioritize solutions generated without relying on RAG-based content, reducing the influence of erroneous data.

We evaluated our method using HumanEval (Chen et al., 2021) and MBPP (Austin et al., 2021). Our approach improves the pass@1 accuracy across all baseline models on both the HumanEval (Chen et al., 2021) and MBPP (Austin et al., 2021) benchmarks by up to 20% compared to the NoRAG method. In comparison to Voyage-Code-2 [1] and BM25 (Robertson et al., 2009), our method demonstrates up to an 8% increase in accuracy on HumanEval and up to a 34% improvement on MBPP. Error analysis on the MBPP dataset, which contains more and complex problems, reveals that assertion errors are reduced significantly, though Name errors are introduced. Additionally, topic analysis on MBPP demonstrate the difficulty of solving some problems e.g., string manipulation when using RAG based on PKG.

In summary, our contribution consists of ①  Programming Knowledge Graph (PKG), a novel representation of code and code-related textual data which represent contents in a DAG to enhance code generation tasks; ② Re-ranking Mechanism, designed to minimize the impact of irrelevant information in RAG methods, by selectively using RAG approaches when needed; ③ Tree Pruning for Semantic Search to remove irrelevant data during the semantic search over the PKG. This approach enhances the accuracy of search results by focusing on meaningful and contextually relevant code blocks.

Our findings demonstrate that the proposed PKG approach along with re-ranker effectively address complex problems while maintaining minimal negative impact on solutions that are already correct without RAG.

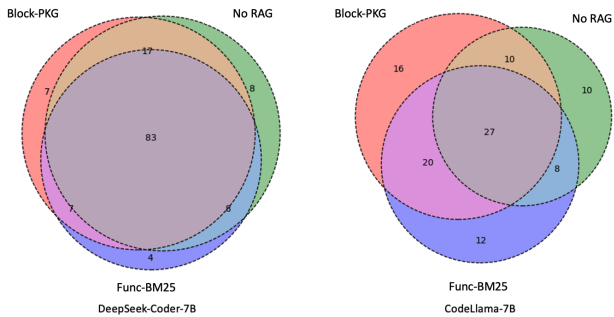

*Figure 1.* This figure illustrates the impact of three approaches – our technique, Programming Knowledge Graph (Block-PKG), Func-BM25, and NoRAG – on solving HumanEval problems using the DeepSeek-Coder-7B (left) and CodeLlama-7B (right) models. Considering CodeLlama-7B, it shows that 16 problems were uniquely solved by the PKG, 12 problems by Func-BM25, and 27 problems were solved by all three approaches.

---

[1]https://blog.voyageai.com/2024/01/23/voyage-code-2-elevate-your-code-retrieval/

## 2. Methodology

Our approach is explained in three distinct steps: (1) PKG Generation, as illustrated in Figure 2, where we describe the process of generating PKG; (2) Information Retrieval from PKG, shown in Figure 3, where we outline the retrieval of relevant information from the PKG; and (3) Solution Re-ranking, where we detail the process of re-ranking the retrieved solutions.

### 2.1. PKG Generation

In this section, we will explain how to generate PKG in 6 steps as explained below.

Step ① Dataset Selection: We generate a PKG from a given set of datasets that contain both text-centric and code-centric content. In our experiments, we have used PythonAlpaca dataset (Petit, 2024) as it consists of conversational question-answers in general python programming problems and Tutorials dataset (Wang et al., 2024), which contains python tutorial text content (Step 1 in Figure 2).

Step ② Block Extraction: This step focuses on extracting structured content blocks, which may consist of function blocks for code-centric data or JSON objects for text-centric content. For code-centric data, an Abstract Syntax Tree (AST) parser, referred to as *FunctionAnalyzer*, is utilized to extract function blocks specifically from the dataset's output sections. For text-centric data, a JSON extraction language model (e.g., Gemma-2 (Team et al., 2024)) is employed. The model is prompted to generate the JSON structure corresponding to the given content. This step is illustrated in Figure 2.

Step ③ Graph Extraction: This section is divided into two parts. First, we detail the process of graph extraction for code-centric data. Subsequently, we describe the methodology for extracting a directed acyclic graph (DAG) from a JSON object.

Step ③.1 Graph Extraction From Function Blocks: In our approach, each code block is represented as a node corresponding to specific programming constructs, such as `if`, `for`, `with`, or `try` blocks. The *FunctionAnalyzer* is responsible for extracting the context-flow graph (CFG) of each function, and subsequently identifying the code blocks, which are represented as individual nodes. Each function consists of three types of nodes: 'function name', 'function implementation', and 'extracted code blocks'. The relationships between these nodes are captured as structural edges in the PKG. Specifically, each function is represented by a 'function name' node, which is connected to a node representing the complete implementation of the function. This implementation node is connected to its corresponding sub-block nodes, reflecting the hierarchical structure of the code (as shown in Step 3 of Figure 2).

Here is the mathematical formulation of the Code Block Extraction process, let $\mathbf{F}$ represent a function. $\mathcal{C}(F)$ be the set of code blocks extracted from $\mathbf{F}$. $G_F = (V_F, E_F)$ represents the graph for the function $F$, where $V_F$ is the set of nodes and $E_F$ is the set of edges representing the relationships between the nodes. The nodes $V_F$ can be defined as:

$$V_F = \{v_{\text{name}}^F, v_{\text{impl}}^F\} \cup \{v_{\text{block}_i}^F | i = 1, 2, \ldots, |\mathcal{C}(F)|\}$$

where $v_{\text{name}}^F$ is the node representing the 'function name', $v_{\text{impl}}^F$ is the node representing the full implementation of function $F$, $v_{\text{block}_i}^F$ represents the $i$-th code block extracted from $F$. The edges $E_F$ capture the hierarchical relationships between the nodes:

$$E_F = \{(v_{\text{name}}^F, v_{\text{impl}}^F) \cup (v_{\text{impl}}^F, v_{\text{block}_i}^F))\} \cup \quad (1)$$
$$\{(v_{\text{block}_j}^F, v_{\text{block}_i}^F) | i, j \in \{1, 2, \ldots, |\mathcal{C}(F)\}|\} \quad (2)$$

The edge $(v_{\text{name}}^F, v_{\text{impl}}^F)$ represents the relationship between the function name and its complete implementation. The edge $(v_{\text{impl}}^F, v_{\text{block}_i}^F)$ represents the relationships between the function implementation and its largest constituent code block and the relations between code blocks are denoted by $(v_{\text{block}_j}^F, v_{\text{block}_i}^F)$. Block-wise retrieval retrieves from $V_{\text{block}}$ while function-wise retrieval only search over $V_{\text{impl}}$ nodes. When we encounter a function call within a retrieved function or code block, we perform a search over the $V_{\text{name}}$ nodes in the knowledge graph. This search allows us to find function calls bodies, enabling us to provide relevant contextual information that makes the retrieved content self-contained.

Step 3.2 Graph Extraction From Json Object: Let $J$ denote a JSON object, which consists of a set of keys and associated values ($J = (key_1, val_1), (key_2, val_2), ..., (key_n, val_n)$).

The JSON object could be represented by $G_J = (V_J, E_J)$, where $V_J$ is the set of nodes $V_J = \{v_1, .., v_m\}$ and $E_J \subseteq V_J \times V_J$ is the set of edges. Each node $v$ represents a path-value pair ($v_i = (path_i, value_i)$) for key $key_i$ in $J$. A path is a concatenated string of keys from the root to the current key $key_i$, forming a unique identifier for each node, and value is the corresponding value in that path. For $key_i$ in $J$, its $path_i$ is shown below, where $key_0, key_j, ..., key_i$ are the keys from the root to the current key ($path_i = key_0\_key_j\_...\_key_i$).

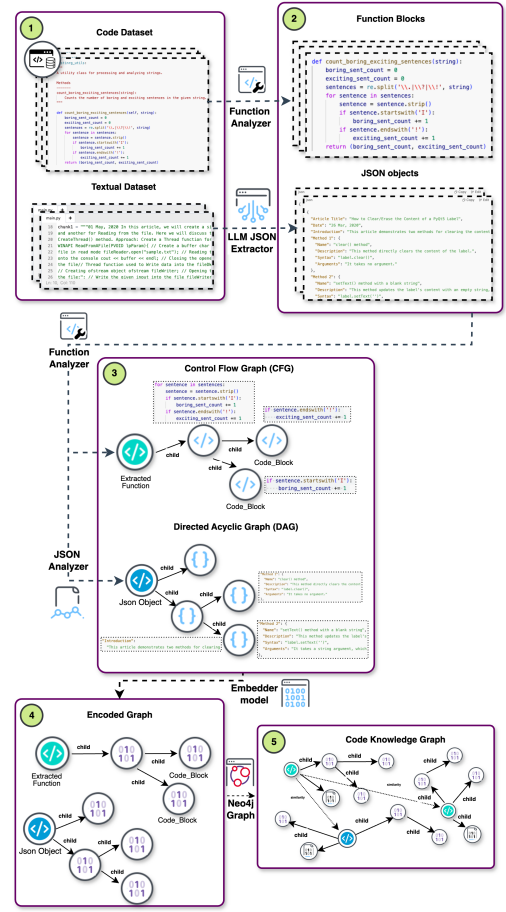

Figure 2. The overview of generating PKG

Value assignment: The value associated with $path_i$ is:

$$value_i = \begin{cases} val_i & \text{if } val_i \text{ is a primitive type (e.g. string, ...)} \\ null & \text{if } val_i \text{ is a JSON object or array} \end{cases} \quad (3)$$

Edge definition: For embedded JSON objects, edges $(v_i, v_j)$ are added between the current node and each of its embedded keys. For lists, edges $(v_i, v_m)$ connect the current node to each list item node.

The final PKG is constructed by aggregating control flow graphs ($G_F$) from code-centric datasets, and directed acyclic graph ($G_J$) from text-centric datasets using Neo4j graph which will be explained in the next steps.

Step 4 Encode PKG: The primary objective of this step is to enable semantic search over the PKG. To achieve this, each node within the graph will be encoded. Previous research, such as the experiments conducted by (Wang et al., 2024), has explored various embedding models for code-RAG methods. Based on these findings, we have selected the VoyageCode2 model[2], which is recognized as one of the most effective embedding models for code representation (Step 4 of Figure 2).

Step ⑤ Neo4j Graph Generation: After extracting all nodes, their corresponding embeddings, and the relationships, we import these data into a Neo4j vector graph. Specifically, we export the nodes and relationships into separate JSON objects and subsequently import them into Neo4j using the APOC plugin. This graph will enable efficient knowledge retrieval through the use of graph indexing and semantic search functionalities.

## 2.2. Retrieval from PKG

To retrieve relevant information for a given query from the PKG, we first obtain the query's embeddings using our embedder model (Step 1 in Figure 3). Let $q$ represent the user query. $\text{Embed}(q) \in \mathbb{R}^d$ be the query's embedding in a $d$-dimensional space, generated by an embedder model $\mathcal{E}$, i.e., $\text{Embed}(q) = \mathcal{E}(q)$. Similarly, for each node $v$ in the PKG, let $\text{Embed}(v) \in \mathbb{R}^d$ represent the embedding of the content of node $v$.

We perform a semantic vector search to identify the node $v_{\text{best}}$ in the PKG that is most similar to the query. This is done by computing the cosine similarity between the query's embedding and each node's embedding (Step 2 in Figure 3):

$$\text{Sim}(q, v) = \frac{\text{Embed}(q) \cdot \text{Embed}(v)}{\|\text{Embed}(q)\| \|\text{Embed}(v)\|} \quad (4)$$

We propose three retrieval approaches on the PKG: block-wise retrieval and function-wise retrieval for $G_F$ and path-value retrieval for $G_J$ to retrieve from different node types. In *Block-wise Retrieval*, retrieval will be performed on the code blocks as a granular retrieval setting, denoted as $v_{block}$, with the results labeled as 'Block-PKG'. This method aims to capture the most relevant context by focusing on related blocks of code within the graph. For *Function-wise Retrieval*, the retrieval will be performed on the implementation nodes, denoted as $v_{impl}$, and the results will be referred to as 'Func-PKG'. The entire function is returned as the relevant context, ensuring that the retrieved information is tightly focused on functional code units. The *Path-value Retrieval* will be performed on the path-value nodes extracted from JSON objects, and the results will be referred to as 'JSON-PKG'. Each retrieved data contains a path points to the value in JSON representations.

At each setting, the node $n_{\text{best}}$ that maximizes this similarity is chosen:

$$n_{\text{best}} = \arg\max_{n \in \mathcal{V}} \text{Sim}(q, n) \quad (5)$$

---
[2] https://docs.voyageai.com/docs/embeddings

Path-value nodes from JSON objects are formulated in a way that the most similar path will be selected using the similarity comparison. For the function blocks ($v_{block}$) and the function implementation ($v_{impl}$) nodes, we refine the selected node $n_{\text{best}}$ by removing branches that are irrelevant to the query (Step 3 in Figure 3). The node $n_{\text{best}}$ is modeled as a Directed Acyclic Graph (DAG) $G_{n_{\text{best}}} = (V_{n_{\text{best}}}, E_{n_{\text{best}}})$, where each node represents a code-block or sub-function, and edges represent child dependencies between them.

For branch pruning, let $G_{n_{\text{best}}}^{-i}$ represent the pruned graph where the $i$-th branch (subgraph) is removed from $G_{n_{\text{best}}}$. We compute the embedding $\text{Embed}\left(G_{n_{\text{best}}}^{-i}\right)$ for each pruned version of the function. The best pruned version $G_{\text{pruned}}$ is selected by maximizing the cosine similarity between the query embedding and the pruned graph embeddings:

$$G_{\text{pruned}} = \arg\max_i \text{Sim}\left(q, G_{n_{\text{best}}}^{-i}\right)$$

Query Augmentation (Step 4 in Figure 3): After identifying the most relevant pruned version of the node, we augment the original query $q$ with the pruned graph content (i.e., $n_{\text{pruned}}$):

$$q_{\text{augmented}} = \text{Augment}(q, n_{\text{pruned}})$$

where $Augment$ is a function that combines the query with the $n_{\text{pruned}}$ content. For instance, as illustrated in Figure 3, if the user's prompt is to generate code that counts the total number of 'boring' sentences starting with 'I', the knowledge graph may initially return a function that counts both 'boring' and 'exciting' sentences. By removing the 'exciting' sentence branch, we refine the function to better align with the query (Step 3 in Figure 3). In the final step, we augment the query with the retrieved function and send it to the model for code generation.

## 2.3. Solution Re-ranking

In our results, we demonstrate that even with access to a PKG or other retrieval sources, the model can still hallucinate when provided with additional information in certain scenarios. This highlights the necessity of incorporating a re-ranking mechanism to effectively select the best solution from multiple approaches. A visual representation of this motivation is provided in Figure 1, which compares the performance of different approaches on the HumanEval benchmark for both CodeLlama-7B and DeepSeek-Coder-7B models. The figure shows that when both BM25 and PKG are applied, 10 problems are solved incorrectly, whereas these same problems are solved correctly without the additional context.

To address this issue, we implemented a simple yet effective re-ranking approach consisting of three key steps. Let $C = \{c_1, c_2, ..., c_n\}$ be the set of initial solution candidates from different approaches (such as Block-PKG, Func-PKG, No RAG,...).

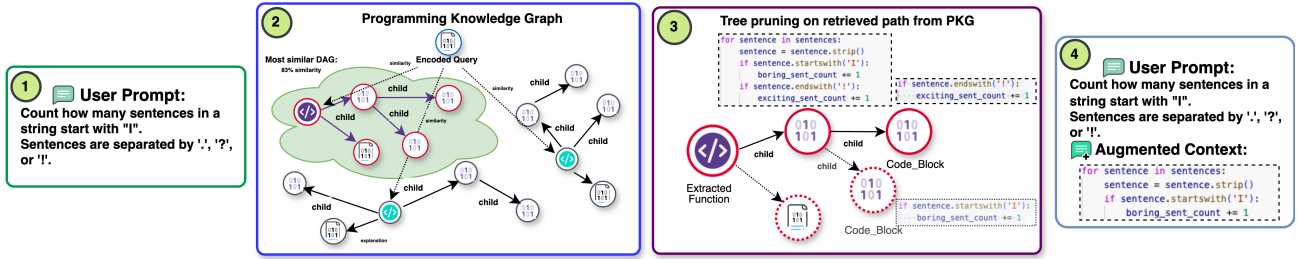

*Figure 3.* Overview of the retrieval process from PKG

First, the solution candidates are passed through AST analysis to filter out those with syntactical errors. Let $A : C \to {0, 1}$ be the AST analysis function where:
$$A(c_i) = \begin{cases} 1 & \text{if } c_i \text{ is syntactically valid} \\ 0 & \text{otherwise} \end{cases}$$

The filtered samples are denoted by $C_A = \{c_i \in C : A(c_i) = 1\}$. In the second step, we execute the remaining candidates($C_A$) to eliminate any solutions containing runtime issues, such as undefined variables. Let $R : C \to {0, 1}$ be the runtime execution function where:
$$R(c_i) = \begin{cases} 1 & \text{if } c_i \text{ executes without errors} \\ 0 & \text{otherwise} \end{cases}$$

The remaining samples after runtime execution phase are denoted by $C_R = \{c_i \in C_A : R(c_i) = 1\}$. Finally, we conduct a semantic similarity check by comparing the embeddings of the remaining candidates with the query embedding. The solution with the highest similarity score is then returned. Let $q$ represent the user query, and $c_i$ denote each candidate selected from the remaining options($C_R$). To calculate the similarity between the query and each candidate, we apply Equation 4, returning the candidate with the highest similarity score. The final selected solution $c^*$ is given by:

$$c^* = \underset{c_i \in C_R}{\arg\max} : sim(q, c_i) \qquad (6)$$

## 3. Related Work

### 3.1. Program Generation Using LLMs

Recent works have demonstrated significant advances in LLM and CLLM code generation (Dubey et al., 2024; Lozhkov et al., 2024; Zhu et al., 2024; Roziere et al., 2023a), evaluated using pass@k metrics (Chen et al., 2021). While models employ various training objectives including code infilling, long context handling (Roziere et al., 2023a), fill-in-the-middle (Li et al., 2023), and instruction fine-tuning (Li et al., 2023; Roziere et al., 2023a; Zhu et al., 2024), our approach uniquely stores and retrieves domain knowledge from a graph structure during generation.

### 3.2. Retrieval Augmented Generation

RAG approaches, widely explored in general text generation (Guu et al., 2020; Lewis et al., 2020; Jiang et al., 2023; Gao et al., 2023), can be categorized into three types: (1) Naive RAG, using a simple retriever for similar content retrieval; (2) Advanced RAG, incorporating query rewriting and solution re-ranking; and (3) Modular RAG, combining multiple strategies to select the best documents. Our framework fits into Modular RAG, utilizing both naive and advanced RAG components.

### 3.3. RAG for Code Generation

RAG's use in code-related tasks remains underexplored (Wang et al., 2024). Previous studies, like (Parvez et al., 2021), focused on smaller models like CodeBERT and GraphCodeBERT for tasks like code summarization and generation, often fine-tuning the retriever. In contrast, our approach applies RAG during inference without fine-tuning. While (Wang et al., 2024) explores LLMs and CLLMs across data sources, they note issues with retrievers and limited model context. Our method improves knowledge representation, enabling more accurate retrieval and reducing hallucinations by prompting models with only relevant content.

## 4. Experimental Setup

**Retrieval Approaches:** We utilized two retrieval methods based on a comparative analysis of various code retrieval models, as described by (Wang et al., 2024). For dense retrieval, we selected the Voyage-Code-2 model, recognized as one of the top-performing dense retrievers for code. Embeddings were obtained through API calls to this model. For sparse retrieval, we employed the BM25 algorithm, implemented using the $rank\_bm25$ Python library[3], which exhibited the strongest performance among sparse retrieval techniques.

---
[3]https://pypi.org/project/rank-bm25/

**Dataset and PKG Generation:** We used the PythonAlpaca dataset (Petit, 2024) as a code-centric data source, which contains 143,000 general Python question-answer pairs. After preprocessing, we extracted 115,000 Python functions from the dataset. This extraction enabled us to construct a PKG comprising 425,058 nodes and 434,518 relations.

We also performed experiments with the the Tutorials dataset (Wang et al., 2024) as a text-centric data source, which contains 76,600 programming tutorial content. After converting them into json representations, pkg contains 288,583 path-value nodes and 287,936 relations. The graphs were generated using Neo4J version 5.20.0, optimized for handling large-scale graphs and supporting semantic search over the stored content.

**Code Generation Models:** We conducted our experiments on four well-known CLLMs: CodeLlama-7B (Roziere et al., 2023b),CodeLlama-13B (Roziere et al., 2023b), StarCoder2-7B (Lozhkov et al., 2024), and DeepSeek-Coder-7B (Zhu et al., 2024). In addition, we tested Llama3.1-8B (Dubey et al., 2024), a general-purpose LLM that has demonstrated strong performance on code generation tasks. All experiments were conducted using a single A100 GPU.

**Evaluation Metric:** To evaluate the accuracy of generated code, we used the pass@1 metric (Chen et al., 2021). Due to resource constraints, we adopted a greedy decoding approach for the pass@1 evaluation, generating a single solution with a temperature setting of $t = 0$ and a token limit of 512 ($max\_new\_tokens = 512$).

**Benchmarks:** In this study, we aim to evaluate the general Python programming skills of both CLLMs and LLMs. To achieve this, we have selected the HumanEval dataset (Chen et al., 2021) and the MBPP benchmark (Austin et al., 2021). These datasets are well-established in the literature and are widely used to assess both problem-solving and reasoning capabilities in Python programming.

## 5. Results

In this section we carry out experiments to answer the following research questions.

**RQ1: Does code-centric PKG improve code generation?**

In this research question, we aim to explore the potential of leveraging graph-based retrieval-augmented methods on code-centric data source to improve code generation task. Specifically, we examine how relevant code retrieved from a PKG built on PythonAlpaca (Petit, 2024) can improve the performance of LLMs and CLLMs in generating accurate code.

Our method retrieves relevant code from the PKG and integrates it into the generation process (See section A.9 in Appendix). We compare this approach to several baselines, detailed in Tables 1 and 2 for HumanEval and MBPP benchmarks. The baselines include: 1) None: No retrieval-augmented generation, 2) BM25: Applied to the entire dataset without pre-processing, 3) VoyageEmb: Embeddings from question-answer pairs for retrieval, 4) Func-BM25: BM25 applied to function-extracted data, 5) Func-PKG: Semantic search over function-related nodes, 6) Block-PKG: Granular semantic search over code blocks for deeper context, 7) Reranked: Re-ranking of candidates from the retrieval methods, and 8) Ideal Re-ranker: An upper bound simulating perfect re-ranking conditions.

As shown in Tables 1 and 2, our approach outperforms NoRAG and other RAG methods across most CLLMs under identical conditions, ensuring a fair comparison with equal data access. However, DeepSeek-Coder shows less improvement in HumanEval, consistent with findings from (Wang et al., 2024), suggesting it may not effectively utilize additional contextual information during training. Figure 1 illustrates the necessity for a re-ranking algorithm: while RAG can solve more problems, it may also degrade some correct solutions by adding external context. Our re-ranking algorithm mitigates this by selecting the best candidate solution, optimizing performance. As shown in the "Reranked" column of Tables 1 and 2, PKG combined with our re-ranker consistently outperforms both benchmarks across all CLLMs and LLM models, significantly improving Pass@1 accuracy for HumanEval and MBPP.

**RQ2: Does text-centric PKG improve code generation?** We explore the potential of using textual data to enhance code generation by building a PKG on a Python-focused subset of the Tutorials dataset from (Wang et al., 2024). The selected content was processed with the Gemma2-9B model (Team et al., 2024), producing hierarchical JSON representations to generate the PKG, as detailed in Section 2.1. We evaluated the JSON-PKG using the HumanEval (Chen et al., 2021) and MBPP (Austin et al., 2021) benchmarks, with results summarized in Table 3 and Table 4. The evaluation shows that JSON-PKG retrieval consistently boosts baseline models' performance, except for DeepSeek-Coder-7B (Zhu et al., 2024). The structured data has a notable positive impact, especially when compared to Llama3.1-8B (Dubey et al., 2024). We hypothesize that general LLMs benefit more from text-centric data as supplementary context than CLLMs. Additionally, while we tested retrieval methods like BM25 and VoyageEmb, JSON-PKG outperformed them in Pass@1 accuracy across both benchmarks. However, when comparing Block-PKG with JSON-PKG, code-centric data still offers greater benefits for code generation tasks, highlighting that code-focused data remains more effective for these specific tasks.

Table 1. Performance of code-centric retrieval-augmented code generation on HumanEval, reported as pass@1. Red cells indicate performance below NoRAG and green cells show scores above NoRAG, with color intensity reflecting significance. "Ideal Reranker" serves as an upper bound for our proposed re-ranker.

| Model | None | BM25 | VoyageEmb | Func-BM25 | Func-PKG | Block-PKG | Reranked | Ideal Reranker |
|---|---|---|---|---|---|---|---|---|
| CodeLlama-7B | 33% | 21% | 42% | 33% | 38% | 40% | **46%** | 56% |
| CodeLlama-13B | 42% | 34% | 45% | 43% | 46% | 47% | **51%** | 63% |
| Llama3.1-8B | 55% | 34% | 50% | 54% | 55% | 61% | **66%** | 75% |
| StarCoder2-7B | 45% | 41% | 53% | 57% | 56% | 59% | **63%** | 72% |
| DeepSeek-Coder-7B | 70% | 44% | 60% | 62% | 69% | 68% | **73%** | 83% |

Table 2. Performance of code-centric retrieval-augmented code generation on MBPP, reported as pass@1. Red cells indicate accuracy below NoRAG, green cells indicate accuracy above, and color intensity reflects significance. "Ideal Reranker" serves as the upper bound for the proposed re-ranker method.

| Model | None | BM25 | VoyageEmb | Func-BM25 | Func-PKG | Block-PKG | Reranked | Ideal Reranker |
|---|---|---|---|---|---|---|---|---|
| CodeLlama-7B | 38% | 27% | 32% | 27% | 44% | 46% | **58%** | 60% |
| CodeLlama-13B | 44% | 36% | 26% | 36% | 40% | 48% | **55%** | 57% |
| Llama3.1-8B | 43% | 38% | 41% | 41% | 46% | 49% | **63%** | 66% |
| StarCoder2-7B | 46% | 25% | 17% | 31% | 29% | 51% | **62%** | 64% |
| DeepSeek-Coder-7B | 56% | 50% | 45% | 47% | 50% | 47% | **65%** | 68% |

Table 3. The performance of PKG on HumanEval, using tutorials data, is reported as pass@1. Red cells indicate accuracy below NoRAG, green cells above, with color intensity reflecting significance.

| Model | None | BM25 | VoyageEmb | JSON-PKG |
|---|---|---|---|---|
| CodeLlama-7B | 33% | 28% | 35% | 36% |
| CodeLlama-13B | 42% | 29% | 43% | 41% |
| Llama3.1-8B | 55% | 47% | 58% | 63% |
| StarCoder2-7B | 45% | 40% | 59% | 61% |
| DeepSeek-Coder-7B | 70% | 60% | 59% | 65% |

Table 4. PKG performance on MBPP using tutorial data, measured by pass@1. Red indicates accuracy below NoRAG, green above, with shading intensity showing significance.

| Model | None | BM25 | VoyageEmb | JSON-PKG |
|---|---|---|---|---|
| CodeLlama-7B | 38% | 29% | 30% | 41% |
| CodeLlama-13B | 44% | 35% | 36% | 45% |
| Llama3.1-8B | 43% | 48% | 51% | 51% |
| StarCoder2-7B | 46% | 49% | 38% | 50% |
| DeepSeek-Coder-7B | 56% | 58% | 51% | 57% |

**RQ3: Which knowledge representation method is most effective in optimizing context retrieval for code generation tasks?**

In this research question, we evaluate RAG performance by exploring three knowledge representations: (1) Question-Answering (Q&A) for entire rows, (2) Function-wise (FW), and (3) Block-wise (BW). We use two retrievers: BM25 (sparse retriever) and Voyage-Code-2 (dense retriever).

We first compare BM25 and Func-BM25 in Tables 1 and 2, showing that low-quality question-answer data (BM25) negatively impacts performance compared to cleaner, function-extracted data (Func-BM25). This highlights the importance of cleaner data for improved code generation and the limited context capacity of models dealing with noisy input. A similar trend appears when comparing VoyageEmb (Voyage-Code-2 on Q&A) with Func-PKG (Voyage-Code-2 on extracted functions), emphasizing the negative effects of irrelevant data in dense retrieval. Finally, the Func-BM25 vs. Func-PKG comparison demonstrates that dense retrieval (i.e., Func-PKG) consistently outperforms sparse retrieval (i.e., Func-BM25) on the same content, underscoring the effectiveness of dense retrievers in capturing nuanced semantic relationships. Finally, comparing Func-PKG to Block-PKG shows that using more granular data, especially at the block level, significantly boosts model accuracy. Block-PKG improves precision by retrieving relevant individual code blocks rather than entire functions. This method prunes irrelevant branches from the DAG associated with selected blocks, ensuring that only the most pertinent context is used. By focusing on finer-grained code structures, Block-PKG delivers superior performance across most models, providing a more targeted and efficient retrieval process.

**RQ4: Which problem topics benefit more from RAG, and which benefit less?**

This research question examines RAG performance across problem categories. Using DeepSeek-Coder-7B, we extract 134 unique categories from MBPP (Austin et al., 2021), a more diverse dataset than HumanEval, and group them into

Table 5. Error Analysis on MBPP for different settings

| Error Type | StarCoder-7B | StarCoder-7B + Block-PKG | CodeLlama-7B | CodeLlama-7B + Block-PKG | DeepSeekCoder-7B | DeepSeekCoder-7B + Block-PKG |
| --- | --- | --- | --- | --- | --- | --- |
| # of AssertionErrors | 198 | 147 | 180 | 162 | 135 | 146 |
| # of NameErrors | 51 | 64 | 138 | 65 | 64 | 78 |
| # of TypeErrors | 11 | 8 | 28 | 37 | 4 | 16 |
| # of SyntaxErrors | 2 | 0 | 0 | 1 | 0 | 0 |
| # of IndentationErrors | 0 | 18 | 0 | 0 | 0 | 0 |
| # of Others | 3 | 7 | 11 | 4 | 5 | 9 |

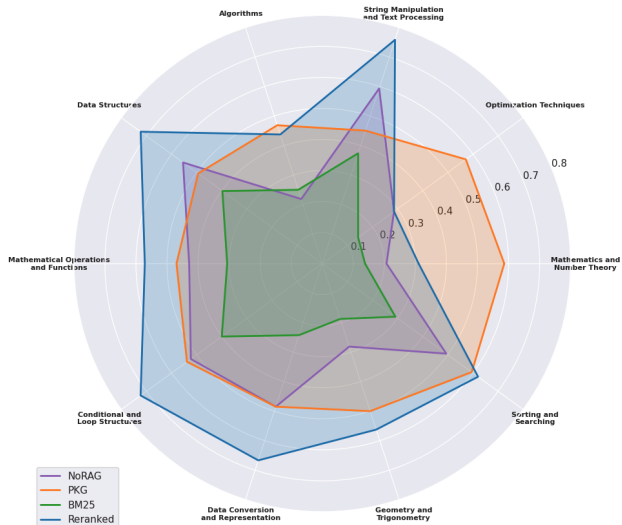

Figure 4. Comparison of different approaches across 10 topics using the MBPP benchmark on StarCoder2-7B

10 broader topics. We then compute pass@1 for each topic to assess RAG effectiveness across domains. Figure 4 shows that PKG consistently outperforms BM25 across all topics and improves accuracy in 7 out of 10 topics compared to NoRAG. However, PKG underperforms on "string manipulation" and "data structures," likely due to the inherent difficulty of string manipulation for token-based generative models (more details in Section A.4). The figure also highlights re-ranking performance. While it struggles with "Optimization Techniques," "Mathematics and Number Theory," and "Algorithms" when applied to Block-PKG, it effectively identifies correct solutions in other topics. Topic analysis for other models is provided in the Appendix.

**RQ5: What types of errors can be reduced or introduced by applying RAG?**

While previous research questions focus on correct solutions in RAG, this one examines incorrect solutions, analyzing error types mitigated or introduced by RAG across StarCoder-7B, CodeLlama, and DeepSeekCoder. Using MBPP execution traces, we assess RAG's impact on error dynamics. As shown in Table 5, StarCoder-7B+PKG re-duces assertion errors by 51 but introduces 18 indentation errors absent in the baseline. CodeLlama7B+PKG reduces name errors by 73 but increases type errors by 9, indicating RAG mitigates assertion errors but can introduce others due to added context. DeepSeekCoder7B, despite accessing the same data, generates more assertion, name, type, and miscellaneous errors, suggesting it struggles to effectively utilize RAG-provided context.

### 5.1. Cost Trade-off

We compare PKG's generation cost with a standard RAG setup using VoyageAI and BM25 retrieval on the PythonAlpaca dataset (Petit, 2024). PKG and VoyageAI share identical encoding times since both use the same embedding model (Voyage-Code2) and dataset (Table 6). Unlike embedding-based RAG methods, PKG requires an additional hour of processing but achieves a 9.4% higher accuracy on average. Neo4j's semantic vector indexing enables efficient graph updates with logarithmic complexity: $O(\log N)$ for nodes and $O(\log M)$ for relationships. Retrieval involves comparing query embeddings with all nodes, resulting in $O(N \cdot d)$ complexity, where $d$ is the embedding dimension. In practice, queries took about 3 seconds each.

Table 6. Time and storage usage for creating RAG data sources on PythonAlpaca (Petit, 2024). Time is in minutes, and storage (last row) is in megabytes (MB).

| Step | PKG | VoyageAI | BM25 |
| --- | --- | --- | --- |
| Python Code Extraction | 3 | - | - |
| Block Extraction | 25 | - | - |
| Encoding | 241 | 240 | 44 |
| Neo4j Graph Generation | 33 | - | - |
| **Overall Time** | **301** | **241** | **44** |
| **Storage Usage (MB)** | **12,530** | **8,440** | **315** |

## 6. Conclusion

We introduced PKG for code generation and evaluated it on HumanEval (Chen et al., 2021) and MBPP (Austin et al., 2021) using both a code-centric dataset (PythonAlpaca (Petit, 2024)) and a text-centric dataset (Tutorials (Wang et al., 2024)). PKG enables fine-grained retrieval of relevant code and related content, while our re-ranker filters out subopti-mal solutions to ensure high-quality outputs. Key findings: (1) PKG-based methods outperform other RAG and non-RAG approaches, (2) LLMs and CLLMs are highly sensitive to irrelevant data, (3) a code re-ranker is crucial for optimizing RAG-based code generation, and (4) problem-topic specificity influences the effectiveness of RAG. Future work should explore improved instruction-tuning techniques and address the lack of dedicated code re-rankers in the literature.

# A. Appendix

In this section, we begin with an ablation study on tree pruning to assess its significance. Next, we conduct experiments to evaluate the effectiveness of the ReRanker steps. Following this, we present an analysis of the token budget required for incorporating additional context in various RAG approaches. We then discuss the retrieval challenges encountered in PKG, followed by a comparison between PKG and related approaches. Finally, we provide a comprehensive analysis of the experimental results from the CodeLlama-7B, StarCoder2, and DeepSeek-Coder-7B models. For each model, we specify the prompt templates used in the experiments to ensure reproducibility and clarity. To facilitate comparison, we also include radar charts that visually depict the accuracy of each model across different problem topics, highlighting their performance on topic-specific tasks.

Additionally, we analyze the distribution of solved and unsolved MBPP problems across various topics, comparing two scenarios: one without RAG (NoRAG) and another using our proposed approach. This comparison highlights the impact of our method on problem-solving effectiveness.

Finally, we present case studies of specific problems where the NoRAG approach fails, but our method succeeds. These examples provide concrete evidence of the advantages of our approach in addressing challenging tasks.

## A.1. Ablation Study on Tree Pruning

In this section, we evaluate the impact of tree pruning through an ablation study on the HumanEval benchmark. To examine its effect, we retrieve only the single most similar node—either a function block or a code block—without applying tree pruning. The retrieved node is then provided as additional context for our baseline models. This approach isolates the influence of tree-pruning on retrieval-augmented code generation, allowing us to assess the effectiveness of tree pruning in refining retrieved context. The results are shown in Table 7.

Table 7. Ablation study on the impact of tree pruning in PKG-based retrieval for HumanEval. We compare results when retrieving content with and without tree pruning.

| Model | Block-PKG | Block-PKG (No Pruning) |
|---|---|---|
| CodeLlama-7B | 40% | 39% |
| CodeLlama-13B | 47% | 44% |
| Llama3.1-8B | 61% | 58% |
| StarCoder2-7B | 59% | 55% |
| DeepSeek-Coder-7B | 68% | 57% |

## A.2. Effectiveness of Re-Ranker Steps

To assess the effectiveness of each step in the Re-Ranker process prior to semantic re-ranking, we conducted an experiment evaluating the quality of generated solutions across both RAG and No-RAG approaches. Specifically, we measured the percentage of solutions containing syntactical errors and runtime execution errors.

Our evaluation focused on MBPP solutions generated by CodeLlama-7B, as reported in Table 8, and HumanEval solutions generated by StarCoder2-7B and Llama3-8B, detailed in Table 9 and Table 10, respectively. The results indicate that both re-ranking steps effectively filter erroneous solutions. Additionally, different models exhibit varying degrees of sensitivity to additional context.

For instance, CodeLlama-7B demonstrates a higher susceptibility to RAG-based approaches, as it generates more syntactical and runtime errors when additional context is appended to its input (see Table 8). In contrast, Llama3-8B appears more robust to context augmentation, maintaining better performance across different RAG approaches.

Table 8. Performance of Re-Ranker steps on MBPP solutions using various RAG approaches with CodeLlama-7B.

| ErrorType | None | BM25-RAG | Func-PKG | Block-PKG |
|---|---|---|---|---|
| **Syntactical Error** | 15% | 33% | 15% | 9% |
| **Runtime Error** | 60% | 74% | 62% | 55% |

Table 9. Performance of Re-Ranker steps on HumanEval solutions using various RAG approaches with StarCoder2-7B.

| ErrorType | None | BM25-RAG | Func-PKG | Block-PKG |
|---|---|---|---|---|
| **Syntactical Error** | 28% | 29% | 31% | 29% |
| **Runtime Error** | 15% | 8% | 5% | 8% |

Table 10. Performance of Re-Ranker steps on HumanEval solutions using various RAG approaches with Llama3-8B.

| ErrorType | None | BM25-RAG | Func-PKG | Block-PKG |
|---|---|---|---|---|
| **Syntactical Error** | 11% | 16% | 13% | 10% |
| **Runtime Error** | 6% | 6% | 3% | 5% |

### A.3. Token Budget Required for Additional Context in Different RAG Approaches

One of the key advantages of retrieving finer-grained contextual information is the ability to provide the model with a reduced token budget, thereby lowering computational inference costs and minimizing monetary expenses, particularly for proprietary models. To evaluate this effect, we compute the average token length of the additional context across different retrieval settings. The results are shown in Table 11

Table 11. The average number of additional tokens in different approaches for the HumanEval benchmark.

| RAG Method | Avg. Tokens (CodeLlama) | Avg. Tokens (DeepSeek) |
|---|---|---|
| **Block-PKG** | **87** | **84** |
| Func-PKG | 188 | 182 |
| BM-25 | 226 | 218 |
| Voyage | 349 | 339 |

### A.4. Challenges in Retrieving Information from PKG

This section discusses scenarios where the PKG fails to retrieve accurate or relevant information. One notable challenge arises when the task requires domain-specific expertise. For example, if the task involves a specialized framework or project-specific code, the PKG must be populated with relevant data from the corresponding domain or project. Failures occur when queries target a graph that lacks such domain knowledge. Addressing this issue necessitates updating the graph with appropriate domain-specific information.

Through topic analysis, we identified that the PKG often struggles with certain problem categories, such as string manipulation. Experimental observations indicate that this challenge stems from the limitations of both the embedder model and the baseline model, which tend to prioritize semantic meaning over structural characteristics of strings.

Example Problem: Write a Python function to convert lowercase characters to uppercase and vice versa, transforming inputs such as "Hello" into "hELLO" and "pYthon" into "PyTHON".

Challenges:

- Embedding Model's Semantic Bias:

    In RAG, the embedder retrieves content primarily based on semantic meaning rather than formatting or structural patterns. For example, it might interpret "Hello" as a greeting, ignoring the case transformation requirement.

- LLM's Tokenization and Semantic Prioritization:

    LLMs tokenize text based on meaning rather than formatting. Consequently, tokens like "Hello" and "hello" are often treated identically, making tasks involving case transformations particularly challenging.

In summary, both RAG retrieval and LLM tokenization emphasize semantic understanding over structural or formatting details, complicating the handling of tasks like string manipulation. This limitation reduces the effectiveness of PKG-based approaches for such problem categories.

## A.5. Comparison with similar works

DocPrompting: Generating Code by Retrieving the Docs

Both DocPrompting (Zhou et al., 2022b) and our PKG-based approach are designed to support code generation across multiple programming languages. However, a fundamental difference lies in the retrieval mechanism. DocPrompting relies on retrieving information from a documentation pool, whereas our method decomposes content into fine-grained semantic nodes within the PKG, integrating both documentation and code. This enables more precise retrieval of relevant knowledge, improving contextual understanding for code generation. Additionally, while our framework has been evaluated using LLMs, DocPrompting has primarily been tested on smaller-scale models, potentially limiting its effectiveness in handling complex programming tasks.

RepoCoder: Repository-Level Code Completion Through Iterative Retrieval and Generation

RepoCoder (Zhang et al., 2023) is specifically designed for repository-level code completion, leveraging an iterative retrieval-generation pipeline to refine the generated code. In contrast, our PKG-based approach is adaptable to a broader range of code-centric and text-centric datasets, making it more versatile in various software development scenarios. Our retrieval mechanism prioritizes extracting the most relevant, fine-grained content from the knowledge base, which helps reduce hallucinations and improve generation accuracy. Unlike RepoCoder's iterative retrieval strategy, which refines results over multiple retrieval cycles, our method ensures that only the most pertinent information is retrieved from the outset, leading to more efficient and precise code generation.

HippoRAG: Neurobiologically Inspired Long-Term Memory for Large Language Models

HippoRAG (Gutiérrez et al., 2024) introduces a neurobiologically inspired approach for integrating long-term memory into large language models, primarily focusing on natural language tasks that require knowledge aggregation from multiple graph-based sources. While both approaches leverage structured knowledge retrieval, our PKG-based framework is specifically designed to enhance code generation by retrieving only the most relevant context while filtering out extraneous information. This targeted retrieval minimizes hallucinations, ensuring that the generated code is both accurate and contextually appropriate. Unlike HippoRAG, which is optimized for general natural language reasoning, our method is tailored to programming-related tasks, emphasizing the seamless integration of documentation and code to improve the quality of generated outputs.

## A.6. CodeLLama7b

### A.6.1. PROMPTS:

The prompts we have used for CodeLlama7B model is provided in Code A.6.1:

```
def codellama_prompt(problem,augmented_data=None):
    if augmented_data:
        prompt = f"""[INST] You are a python programmer. Solve the following problem:\n{
            problem} \n\nThe following code might be helpful:\n{augmented_data}\nIf helper
             section is useful, integrate their logic directly into the body of the main
             function, otherwise just ignore them. You MUST write your solution between [
            PYTHON] and [/PYTHON]. Your solution MUST be executable.[/INST]"""
        return prompt
    else:
        prompt = f"""[INST] You are a python programmer. Solve the following problem:\n{
            problem} \n\nPlease write the python solution inside [PYTHON] and [/PYTHON]
             tags.\n[/INST]"
        """
        return prompt
```

### A.6.2. TOPIC-SPECIFIC APPROACH COMPARISON:

Figure 5 presents the Pass@1 accuracy for each method—NoRAG, PKG, BM25, and the re-ranked approach—across various programming topics. Similar to the performance observed with the StarCoder2-7B model, the re-ranker struggles to correctly prioritize solutions in the 'Optimization Techniques,' 'Mathematics,' and 'Algorithm' categories. However, in other topic areas, the re-ranker demonstrates superior performance compared to the other methods. Notably, for this model,

PKG achieves higher accuracy across most topics, with the exception of 'String Manipulation' and 'Data Structures,' where it is outperformed by other approaches.

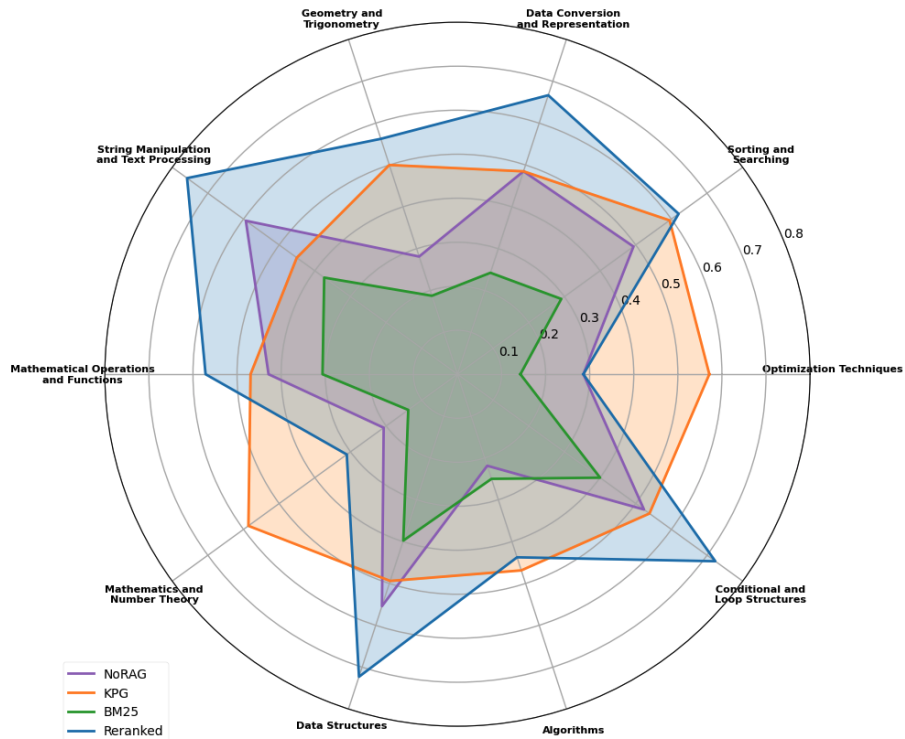

*Figure 5.* Comparison of different approaches across 10 topics using the MBPP benchmark on CodeLlama-7B.

### A.6.3. TOPIC-BASED ACCURACY DISTRIBUTION

Figure 6 illustrates the distribution of MBPP problems on a two-dimensional plot, where the embedding dimensions have been reduced to two for visualization purposes. The different problem topics are represented by distinct shapes, while the correctness of the solutions is indicated by color. Problems that were solved incorrectly are shown in orange, and those solved correctly are shown in green. The legend for each topic separates the total number of correct solutions from the incorrect ones using a slash ("/"). Figure 7 shows the distribution of correct and incorrect problems when we apply our approach.

## A.7. StarCoder2-7B

### A.7.1. PROMPTS:

The prompts we have used for StarCoder2-7B model is provided in Code A.7.1:

```
def starcoder_prompt(problem,augmented_data=None):
    if augmented_data:
        prompt = f"""### Instruction
        You are a python programmer. Solve the following problem:\n{problem} \n\n The
            following code might be helpful:\n{augmented_data}\n. If they are useful,
            integrate their logic directly into the body of the main function, otherwise
            just ignore them.\n
        ### Response
        """
        return prompt
    else:
        prompt = f"""### Instruction
```

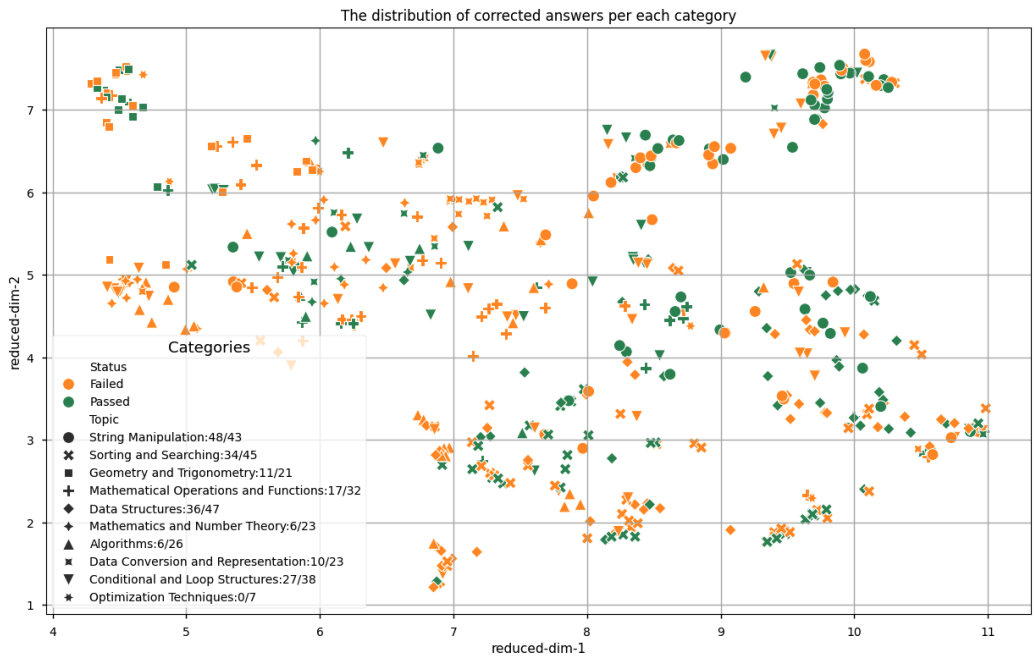

*Figure 6.* The distribution of MBPP solutions on each topic in NoRAG setting.

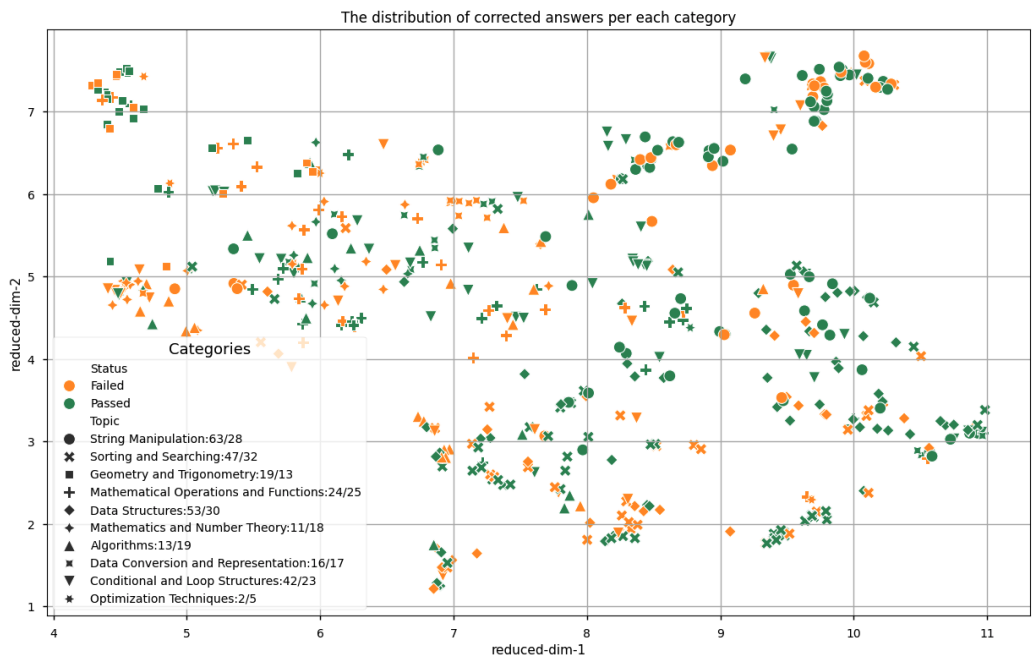

*Figure 7.* The distribution of MBPP solutions on each topic using our re-ranker.

```
11          You are a python programmer. Solve the following problem:\n{problem} \n\n
12          ### Response
13          """
14      return prompt
```

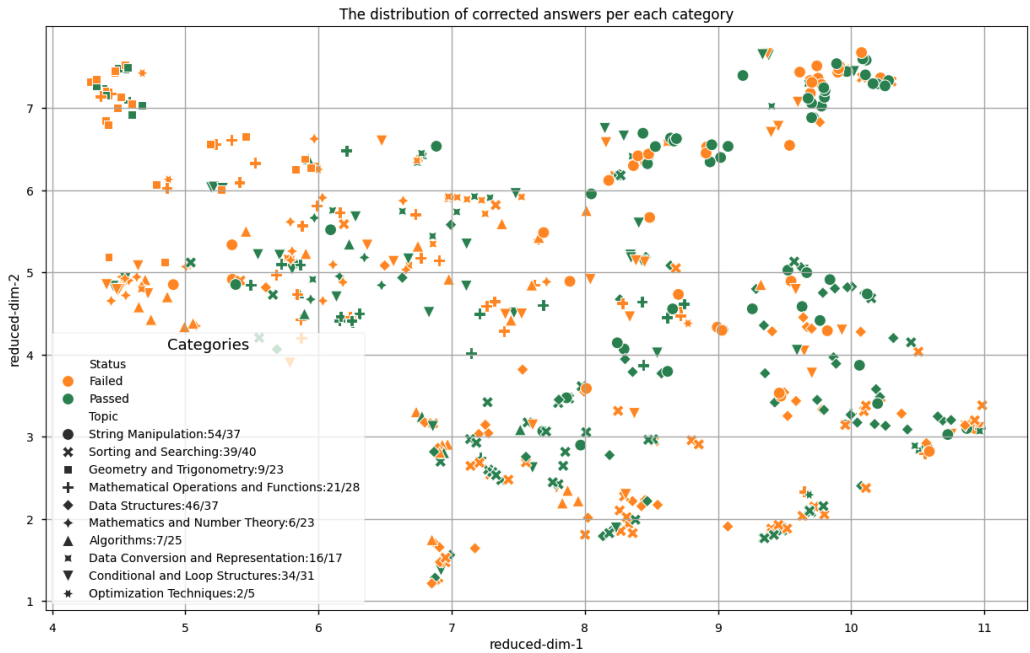

Figure 8. The distribution of MBPP solutions on each topic without RAG.

### A.7.2. TOPIC-BASED ACCURACY DISTRIBUTION

Figure 8 presents the distribution of MBPP problems on a two-dimensional plot, with the embedding dimensions reduced for visualization. Each problem topic is represented by a unique shape, while solution correctness is color-coded. Problems incorrectly solved by StarCoder2-7B are highlighted in orange, whereas correctly solved problems are shown in green. The legend for each topic indicates the total number of correct versus incorrect solutions using a "correct/incorrect" format.

Additionally, Figure 9 visualizes the same distribution but reflects the accuracy after applying our proposed approach, showcasing improvements in solution correctness across topics.

### A.8. DeepSeek-Coder-7B

#### A.8.1. PROMPTS:

The prompts we have used for DeepSeek-Coder-7B model is provided in Code A.8.1:

```
1
2 def deepseek_prompt(problem,augmented_data=None):
3     if augmented_data:
4         prompt = f"""[INST] You are a python programmer. Solve the following problem:\n{
              problem} \n\n The following code might be helpful:\n{augmented_data}\n.If they
               are useful, integrate their logic directly into the body of the main function
              , otherwise just ignore them.\n[/INST]"""
5         return prompt
6     else:
7         prompt = f"""[INST] You are a python programmer.  Solve the following problem: \n
              {problem} \n\n[/INST]"""
8         return prompt
```

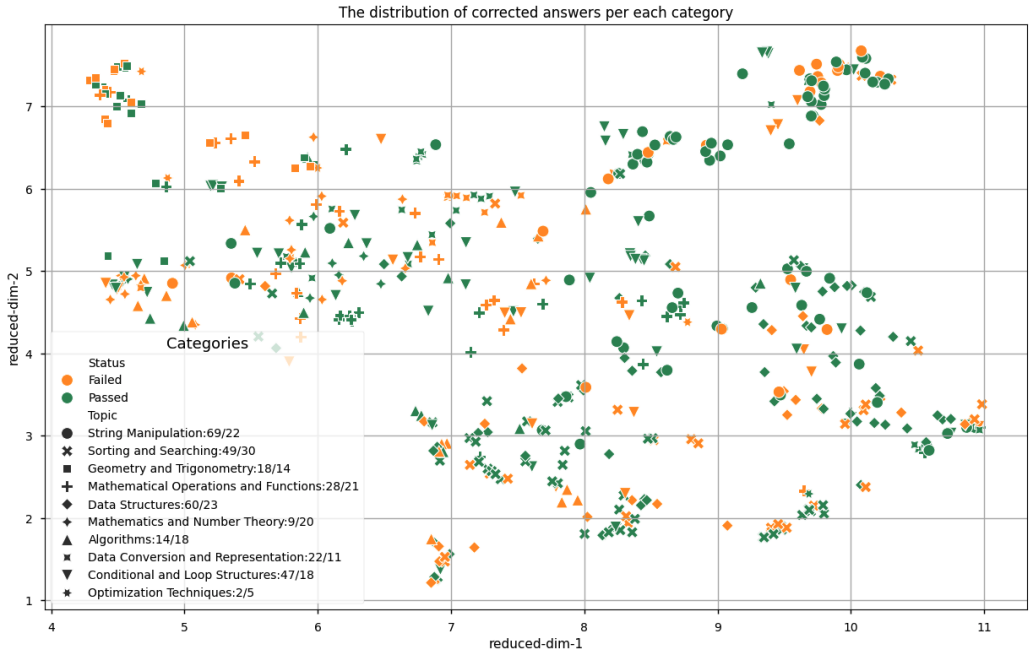

*Figure 9.* The distribution of MBPP solutions on each topic using our proposed re-ranker.

### A.8.2. TOPIC-SPECIFIC APPROACH COMPARISON

Figure 10 illustrates the Pass@1 accuracy for each evaluation method: NoRAG, PKG, BM25, and the re-ranked approach, across a range of programming topics. The performance trends observed with the DeepSeek-Coder-7B model are echoed here. Specifically, the re-ranking method shows difficulty in accurately prioritizing solutions within the categories of 'Optimization Techniques,' 'Mathematics,' and 'Algorithms.' Despite these challenges, the re-ranked approach excels in other topic areas, demonstrating superior performance compared to the other methods.

Notably, the PKG method achieves higher accuracy across most topics evaluated. However, it does face competition in the 'String Manipulation' and 'Data Structures' categories, where it is outperformed by NoRAG approach. We have observed the same behaviour for the previous models.

### A.8.3. TOPIC-BASED ACCURACY DISTRIBUTION

Figure 11 displays the distribution of problems from the MBPP dataset in a two-dimensional plot, achieved by reducing the embedding dimensions for improved visualization. Each distinct shape in the plot corresponds to a specific problem topic, while the correctness of the solutions is indicated by color coding. Problems that were solved incorrectly are represented in orange, whereas those that were solved correctly are shown in green. The legend accompanying each topic delineates the total number of correct solutions from the incorrect ones, separated with a slash ("/").

In addition, Figure 12 presents a similar distribution of problems, highlighting the outcomes after applying our novel approach. This figure further distinguishes between correct and incorrect solutions, allowing for a comparative analysis of the effectiveness of our method.

### A.9. Examples:

In this section, we present two selected samples from the HumanEval benchmark. We provide the responses generated by StarCoder-2-7B and DeepSeek-Coder-7B models. Each model's output is displayed in two scenarios: first, without using RAG, and second, utilizing our PKG approach. These examples illustrate how incorporating additional context can enhance the models' ability to solve complex problems more effectively.

HumanEval problem 159, solved by Starcoder2-7B without RAG (Failed):

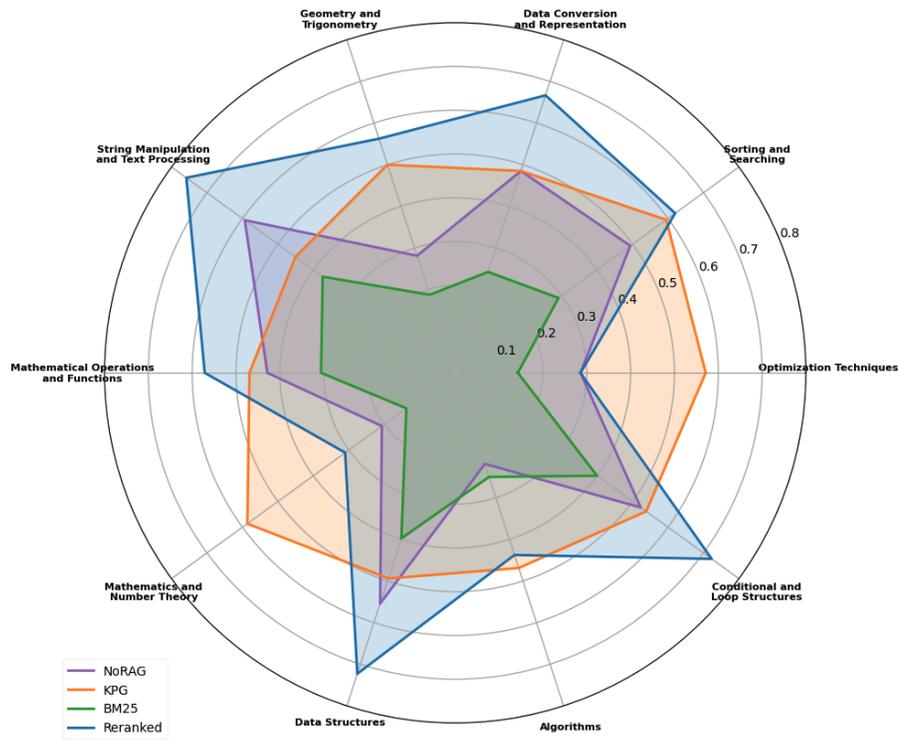

*Figure 10.* Comparison of different approaches across 10 topics using the MBPP benchmark on DeepSeek-Coder-7B

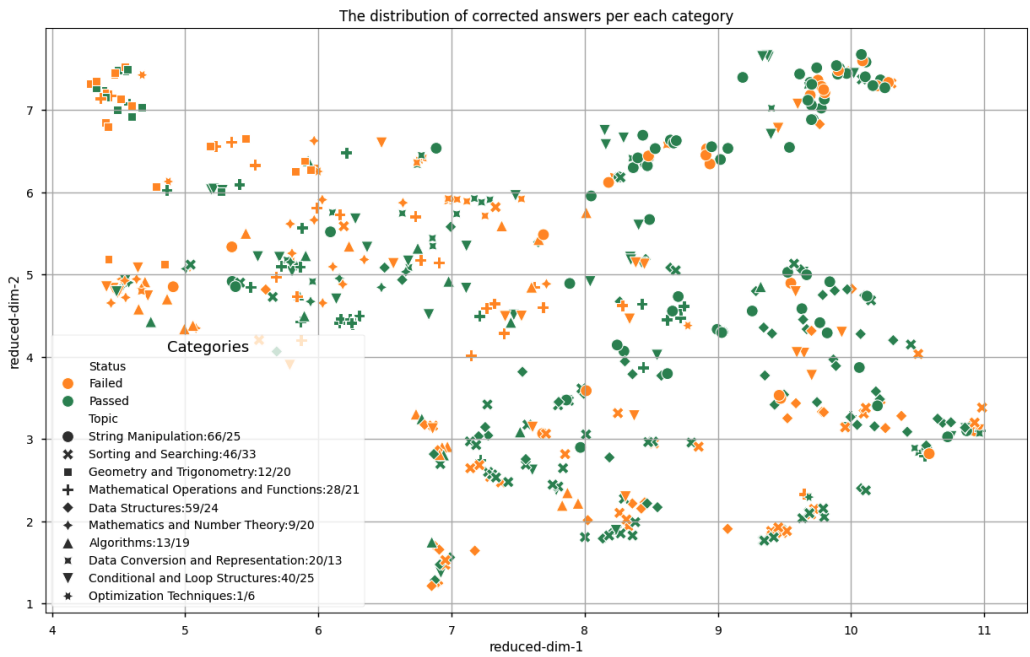

*Figure 11.* The distribution of MBPP solutions on each topic in NoRAG setting.

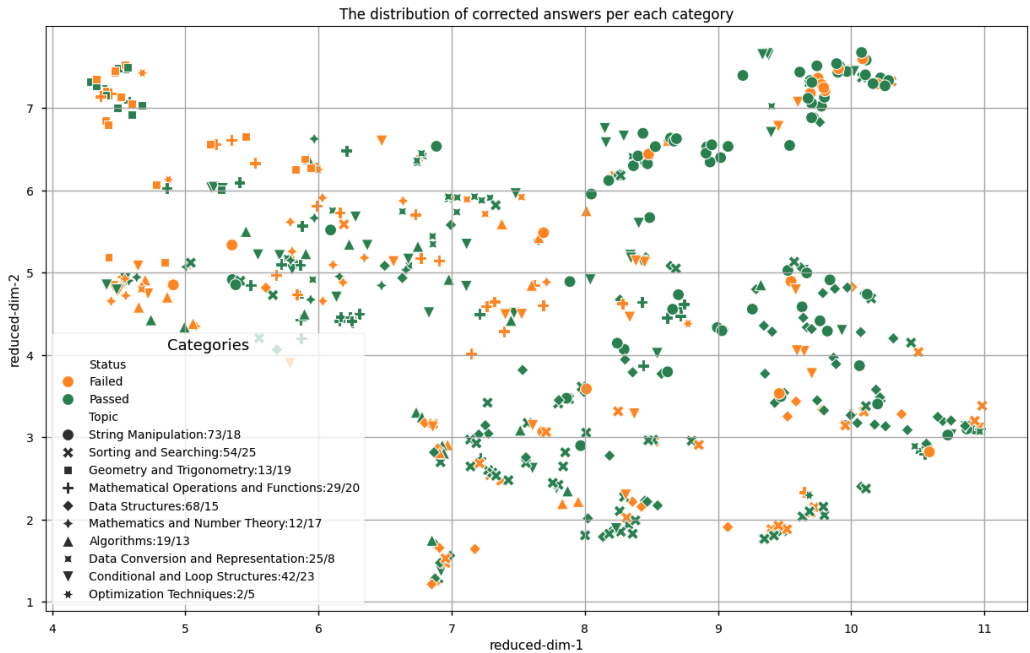

*Figure 12.* The distribution of MBPP solutions on each topic using our reranker.

```
def eat(number, need, remaining):
    """
    You're a hungry rabbit, and you already have eaten a certain number of carrots,
    but now you need to eat more carrots to complete the day's meals.
    you should return an array of [ total number of eaten carrots after your meals,
                                    the number of carrots left after your meals ]
    if there are not enough remaining carrots, you will eat all remaining carrots, but
        will still be hungry.

    Example:
    * eat(5, 6, 10) -> [11, 4]
    * eat(4, 8, 9) -> [12, 1]
    * eat(1, 10, 10) -> [11, 0]
    * eat(2, 11, 5) -> [7, 0]

    Variables:
    @number : integer
        the number of carrots that you have eaten.
    @need : integer
        the number of carrots that you need to eat.
    @remaining : integer
        the number of remaining carrots thet exist in stock

    Constrain:
    * 0 <= number <= 1000
    * 0 <= need <= 1000
    * 0 <= remaining <= 1000

    Have fun :)
    """
    return [number + need, remaining - need]
```

HumanEval problem 159, solved by Starcoder2-7B using PKG (Passed):

```python
### Instruction
#You are a python expert programmer. Implement a python function to #solve the following
    problem:

def eat(number, need, remaining):
    """
    You're a hungry rabbit, and you already have eaten a certain number of carrots,
    but now you need to eat more carrots to complete the day's meals.
    you should return an array of [ total number of eaten carrots after your meals,
                                    the number of carrots left after your meals ]
    if there are not enough remaining carrots, you will eat all remaining carrots, but
        will still be hungry.

    Example:
    * eat(5, 6, 10) -> [11, 4]
    * eat(4, 8, 9) -> [12, 1]
    * eat(1, 10, 10) -> [11, 0]
    * eat(2, 11, 5) -> [7, 0]

    Variables:
    @number : integer
        the number of carrots that you have eaten.
    @need : integer
        the number of carrots that you need to eat.
    @remaining : integer
        the number of remaining carrots thet exist in stock

    Constrain:
    * 0 <= number <= 1000
    * 0 <= need <= 1000
    * 0 <= remaining <= 1000

    Have fun :)
    """

#The following helper code might be helpful:
#helper code 1:
    postEatWeight = totalWeight - amountEaten
    combinedWeight = postEatWeight / 7
    largerBoxPostEat = combinedWeight * 4
    smallerBoxPostEat = combinedWeight * 3
    largerBox = largerBoxPostEat + amountEaten
    smallerBox = smallerBoxPostEat
    return (round(largerBox), round(smallerBox))
#End of helper section.
#Write your solution inside [PYTHON] and [/PYTHON] tags. The solution #should be
    executable and do NOT call helper functions. If they are #useful, integrate their
    logic directly into the body of the main #function, otherwise just ignore them.
### Response
#Here is the Python function to solve the problem:

def eat(number, need, remaining):
    if remaining >= need:
        return [number + need, remaining - need]
    else:
        return [number + remaining, 0]
```

HumanEval problem 90, solved by DeepSeek-Coder2 without RAG (Failed):

```python
def next_smallest(lst):
    """
    You are given a list of integers.
```

```
     Write a function next_smallest() that returns the 2nd smallest element of the list.
     Return None if there is no such element.

     next_smallest([1, 2, 3, 4, 5]) == 2
     next_smallest([5, 1, 4, 3, 2]) == 2
     next_smallest([]) == None
     next_smallest([1, 1]) == None
     """
     if len(lst) < 2:
         return None
     else:
         lst.sort()
         return lst[1]
```

HumanEval problem 90, solved by DeepSeek-Coder2 using PKG (Passed):

```
### Instruction:
#You are an expert python programmer. Solve the following PROBLEM in #python:

def next_smallest(lst):
    """
    You are given a list of integers.
    Write a function next_smallest() that returns the 2nd smallest element of the list.
    Return None if there is no such element.

    next_smallest([1, 2, 3, 4, 5]) == 2
    next_smallest([5, 1, 4, 3, 2]) == 2
    next_smallest([]) == None
    next_smallest([1, 1]) == None
    """

 #The following helper code might be helpful:
 #helper code 1:
    if len(lst) < 2:
        return None
    smallest = float('inf')
    next_smallest = float('inf')
    for i in range(0, len(lst)):
        if lst[i] < smallest:
            next_smallest = smallest
            smallest = lst[i]
        elif lst[i] > smallest and lst[i] < next_smallest:
            next_smallest = lst[i]
    return None if next_smallest == float('inf') else next_smallest

#End of helper section.
#Please write the python solution inside [PYTHON] and [/PYTHON] tags.
### Response:

def next_smallest(lst):
    if len(lst) < 2:
        return None
    smallest = second_smallest = float('inf')
    for element in lst:
        if element < smallest:
            second_smallest = smallest
            smallest = element
        elif smallest < element < second_smallest:
            second_smallest = element
    return None if second_smallest == float('inf') else second_smallest
```